\documentclass[fleqn,usenatbib]{mnras}
\usepackage{newtxtext,newtxmath}

\usepackage[T1]{fontenc}
\usepackage{ae,aecompl}

\usepackage{graphicx}	
\usepackage{amsmath}	
\usepackage{amssymb}	
\usepackage[flushleft]{threeparttable}
\usepackage{color}
\usepackage{natbib}
\usepackage{soul}

\newcommand\solarmass{$M_\odot$}
\newcommand\cygx{Cyg~X-1}
\newcommand\ob{Cyg~OB3}
\newcommand\kms{km~s$^{-1}$}
\newcommand\er{$\pm$}
\newcommand\vpec{$v\sub{pec}$}
\newcommand\vrel{$v\sub{rel}$}
\newcommand\vr{$v\sub{r}$}
\newcommand\us{$U\sub{s}$}
\newcommand\vs{$V\sub{s}$}
\newcommand\ws{$W\sub{s}$}
\newcommand\gaia{\textit{Gaia}}

\newcommand{\sub}[1]{\ensuremath{_{\textrm{#1}}}}

\title[Cyg X-1 and OB3]{Kinematic study of the association Cyg OB3 with \textit{Gaia} DR2}
\author[Rao et al.]{Anjali Rao$^{1} $\thanks{E-mail: a.rao@soton.ac.uk},
Poshak Gandhi$^{1},$
Christian Knigge$^{1},$
John A. Paice$^{1,2},$
\newauthor
Nathan W. C. Leigh$^{3,4},$
and Douglas Boubert$^{5}$
\\
$^{1}$Department of Physics \& Astronomy, University of Southampton, Highfield, Southampton SO17 1BJ, UK\\
$^{2}$Inter-University Centre for Astronomy and Astrophysics, Pune, Maharashtra 411007, India\\
$^{3}$Departamento de Astronom\'ia, Facultad de Ciencias F\'isicas y Matem\'aticas,
Universidad de Concepci\'on, Concepci\'on, Chile\\
$^{4}$Department of Astrophysics, American Museum of Natural History, New York, NY 10024, USA\\
$^{5}$Magdalen College, University of Oxford, High Street, Oxford OX1 4AU, UK\\
}

\date{Accepted XXX. Received YYY; in original form ZZZ}

\pubyear{2020}

\begin{document}
\label{firstpage}
\pagerange{\pageref{firstpage}--\pageref{lastpage}}
\maketitle

\begin{abstract}
We study the stellar kinematic properties and spatial distribution of the association \ob{} using precise astrometric data from \textit{Gaia} DR2. All known O- and B-type stars in \ob{} region with positions, parallaxes and proper motions available are included, comprising a total of 41 stars. The majority of stars are found to be concentrated at a heliocentric distance of 2.0\er{}0.3 kpc. The mean peculiar velocity of the sample after removing Galactic rotation and solar motion is $\sim$22 \kms{}, dominated by the velocity component towards the Galactic center. The relative position and velocity of the black hole X-ray binary \cygx{} with respect to the association suggest that \ob{} is most likely its parent association. The peculiar kinematic properties of some of the stars are revealed and are suggestive of past stellar encounters. The sample includes a previously known runaway star HD~227018, and its high peculiar velocity of $\sim$50 \kms{} is confirmed with \textit{Gaia}. We estimated the velocities of stars relative to the association and the star HD~225577 exhibits peculiar velocity smaller than its velocity relative to the association. The star has lower value of proper motion than the rest of the sample. The results suggest a slowly expanding nature of the association, which is supported by the small relative speeds $<$~20 \kms{} with respect to the association for a majority of the sample stars.

\end{abstract}




\section{Introduction}

OB associations are regions in the sky with higher densities of O- and B-type stars relative to the field \citep{ambartsumian47}. These systems are identified with sites of the most recent star formation regions in the Galaxy and usually host the youngest and most massive stars \citep[see review by][and references therein]{blaauw64}. It has long been known that associations cannot be bound by their own self-gravity and tend to form unbound stellar aggregates spanning tens of parsecs \citep{ambartsumian47, ambartsumian55, melnik17}. Many OB associations are known in the Galaxy and catalogued in papers such as \citet{hum78}, ~\citet{rup81}, \citet{hump-mcelroy84}, \citet{blaha-hum-89}, \citet{garmany92}, and \citet{melnik95}. OB associations have historically been of interest for several reasons, for instance in the investigation of the motions of loosely bound or unbound systems in the Galactic potential \citep[e.g.][]{lindblad41, lindblad42}, the dynamical interactions of the member stars \citep[e.g][]{wright18}, and the formation of stars in giant molecular clouds \citep{clark05}. Recently, \textit{Gaia} DR2 has significantly contributed to the study of OB associations e.g. \citet{cantat}, \cite{ber19}, \citet{wright19} and \citet{melnik20}.

The membership of associations has been assigned and revised by a number of authors, as historical observational advances have been made. \cite{hum78} provided a list of stars in known associations, but only included stars for which a measurement of the photometric distance was possible based on the information from MK spectral types and UBV photometry. \citet{rup81} also presented a catalogue of star clusters and associations. \cite{hump-mcelroy84} expanded the catalogue of \citet{hum78}, and it was further updated by \cite{blaha-hum-89}, where the authors catalogued a total of 2263 stars in 91 associations and clusters and another set of 2549 stars in the field. Later, \cite{garmany92} also provided a list of member stars in associations and computed their distances using the method of cluster fitting of B-type main sequence stars.
In the classification scheme discussed in \cite{hum78}, \cite{hump-mcelroy84}, \cite{blaha-hum-89} and \cite{garmany92}, membership assignment is based on a combination of position on the sky, photometric distance and radial velocity.

\citet{melnik95} adopted a different method, compared to the previous authors, to determine the minimum unitary structures among the stellar associations by implementing an algorithm for cluster analysis. They found the groups and their members and provided a list of new associations in the Galaxy. Clustering depends on the choice of length scale and groups begin to merge for larger length scales. In this scenario, although the presence of different groups and associations is ascertained from the results of their cluster analysis using an independent method, their boundaries are not well defined. This, consequently, affects the membership assignment and identification of parent associations for those stars located at the association boundaries.

\begin{figure}
	\centering
	\includegraphics[height=7.5cm, width=7.5cm]{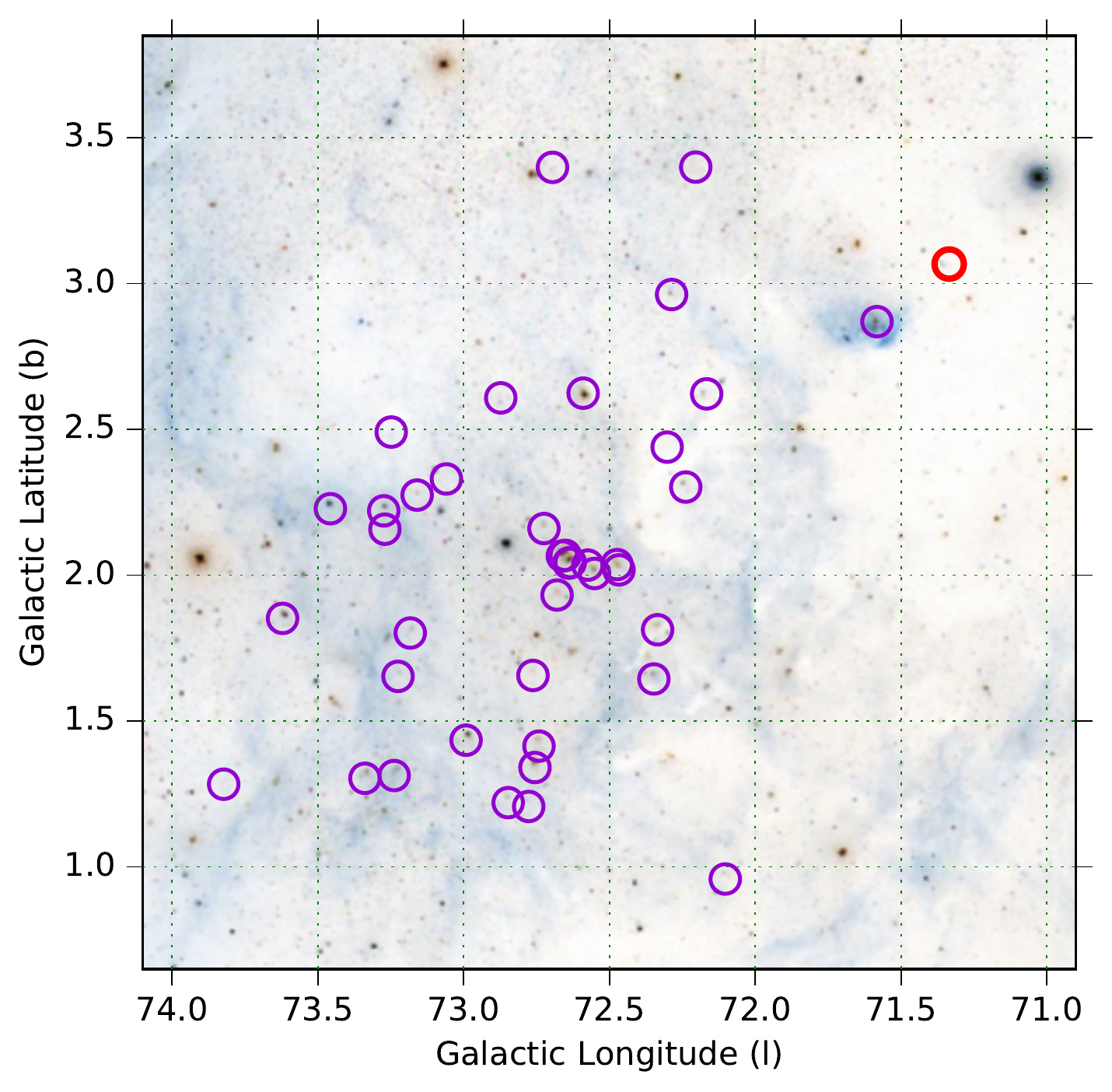}

\caption{Image obtained with DSS2 in the $R$-band shows the \ob{} region. The positions of the O- and B-type stars studied in this work are encircled. \cygx{} is encircled in red. There are a total of 41 stars shown in the image and given in Table~\ref{tab1}.}
	\label{cygregion}
\end{figure}

Here, we investigate the properties of the well known association \ob{}. We primarily focus on the distances, proper motions and three-dimensional velocity components of the members. We study a particular case of a confirmed black hole X-ray binary \cygx{} \citep[otherwise known as HD 226868,][]{bolton72,webster72} located at the association boundary (Fig.~\ref{cygregion}). \cygx{} is one of the brightest X-ray binaries in the X-ray sky and is composed of a black hole of 14.8$\pm$1.0 {\solarmass} \citep{orosz11} and a companion of 19.2$\pm$1.9 {\solarmass} \citep{orosz11} of spectral type O9.7Iab \citep{walborn73}. \cygx{} has been included in the catalogs of \citet{hum78}, \citet{garmany92}, \citet{rup81} and \citet{blaha-hum-89} as a member of \ob{}. The cluster analysis algorithm implemented by \citet{melnik95} divided the association \ob{} into two subgroups for a length scale of 15 pc viz. Cyg~3A and Cyg~3B, however \cygx{} was not included in either of the subgroups and remained as a field star.

\citet{mirabel-rodriguez-2003} studied the connection between \cygx{} and the association \ob{} using the astrometry from the Hipparcos mission. The authors suggested that the parent association of \cygx{} is indeed \ob{}, and showed that the binary is located at the same distance as the other stars in \ob{} and exhibits a proper motion similar to that of the association. The authors also showed that the peculiar velocity of \cygx{} is very small (9 $\pm$ 2 km s$^{-1}$) with respect to the association, and is consistent with the expected random velocities of stars in expanding associations. These results indicate that \cygx{} is most likely a member of \ob{}. This was later confirmed with precise 
measurement of the trigonometric parallax and the proper motion of \cygx{} using VLBI radio observations \citep{reid11}. In this paper, we revisit the proposed connection between \cygx{} and \ob{} using the astrometric data from \textit{Gaia} DR2, which provides trigonometric parallaxes and proper motions for over 1 billion stars in the Galaxy with the best accuracy and precision available to date \citep{gaiadr2}. 

We define the sample of stars and their filtering criteria in the next section. The results are discussed in detail in Section \ref{distancefinding}. Our study of proper motions is discussed in Section \ref{secpm}. Error analysis reveals that the parallax uncertainties are the most significant contributor to the uncertainty when determining the three-dimensional distances between stars (see Section \ref{vpecfinding}). In addition, we find the peculiar velocities and relative velocities of individual stars in the sample, where radial velocity measurements were available from the literature, with respect to the median velocity of the association. A comparison of the two velocities turns out to be an efficient method to filter out stars with peculiar kinematic behavior from the main body of the association. The results are discussed in Section \ref{discuss} and individual cases are discussed in detail.

\section{Observations}

\subsection{Sample of Stars}

The list of associations published by \citet{hum78} was restricted to only those stars for which photometric distances were available and included a total of 25 O- and B-type stars in \ob{}. The list provided by \cite{garmany92} includes a larger sample of 37 stars in the association. According to the lists by \cite{hum78} and \cite{garmany92}, the association \ob{} is located between Cyg~OB1 and Cyg~OB5 within Galactic coordinates of 71$^\circ \leq l \leq$ 74$^\circ$, and 0.9$^\circ \leq b \leq$ 3.5$^\circ$. Instead of restricting to the list of stars given in the above catalogues, we have searched the entire region within the above mentioned Galactic coordinates, and queried the \textit{Gaia} DR2 archive for sources with 5-parameter astrometric solutions available with position, parallax and proper motion ({\tt astrometric\_params\_solved = 31}). Since the faintest OB star in the list of \cite{garmany92} is of magnitude $V$=10.32, stars fainter than $G$=10.30 magnitude are excluded in the query. The relation between $V$ band and Gaia's $G$ band is obtained from Gaia data release documentation\footnote{https://gea.esac.esa.int/archive/documentation/GDR2/Data\_processing/
chap\_cu5pho/sec\_cu5pho\_calibr/ssec\_cu5pho\_PhotTransf.html}. This photometric selection restricts the sample to contain stars of early B-type, in fact all the stars in \cite{garmany92} are of spectral type B5 or earlier.
\ob{} adjoins Cyg~OB1 and there is no clear boundary between the two associations. 
Therefore the sample of stars in \ob{} is expected to be contaminated by the stars of Cyg~OB1 and vice versa. In order to reduce the contamination of stars from Cyg~OB1, the query rejects the sources located at a distance, obtained from parallax inversion, smaller than 1.4 kpc. This distance threshold is chosen because it is the average distance for Cyg~OB1 reported in \cite{melnik17}. We also removed all the stars from the sample at distances $\gtrsim$3 kpc to remove distant non-members. Since the sample has not been filtered based on proper motion or other kinematic properties, the sample may represent more than one kinematic substructure as has been observed in other OB associations e.g., \cite{cantat}.

The resultant list of sources from the query is cross-matched against {\tt SIMBAD} using a 2 arc second search radius, and the spectral types of matched stars are extracted. A total of 6 stars were removed where more than one potential cross-match was found in order to avoid source confusion while assigning {\tt SIMBAD} objects to \textit{Gaia} counterparts. A comparison of magnitudes was not helpful in the identification of sources in these six cases. Four of these systems are binary. Since astrometric solutions of binary systems are not reliable, these were removed from the sample. Stars with spectral type O and B are shortlisted in the sample. All the stars are bright with $G$-band magnitudes in the range 7.19-10.30 with a median value of 9.16.

\subsection{Filtering Criteria}

We followed the three filtering criteria recommended by the \textit{Gaia} team described in \citet{linde}, and also in equations (1-3) of \citet{arenou18}, to extract good astrometric solutions. We also filtered the solutions by re-normalized unit weight error (RUWE) \footnote{See the technical note GAIA-C3-TN-LU-LL-124-01 at https://www.cosmos.esa.int/web/gaia/dr2-known-issues}, defined as $u_{\mathrm{norm}}$ = $u/u_0(G,C)$, where $u$ =  {\tt (astrometric\_chi2\_al/astrometric\_n\_good\_obs\_al} - 5)$^{1/2}$. RUWE is a quality metric introduced after the release of DR2, based on analysis of reliability of the astrometric solutions by the \textit{Gaia} team. Solutions with RUWE $\geq$ 1.4 were rejected.
Finally, another star HD 227573 was removed from the sample, as it has a fractional uncertainty on its parallax $>$ 1. A number of flags viz. {\tt astrometric\_gof\_al}, {\tt astrometric\_chi2\_al}, {\tt astrometric\_excess\_noise} and {\tt astrometric\_excess\_noise\_sig} often used to interpret the goodness of an astrometric fit suggested a bad fit for this star. A total of 4 stars were discarded by these filters. There are 22 stars in common between the list provided by \citet{garmany92} and our sample. Table~\ref{tab1} gives a list of all these stars. The names of the sources obtained from {\tt SIMBAD} and their \textit{Gaia} DR2 source IDs are provided in the second and third columns, respectively. The remaining columns list the DR2 values including RA and Dec coordinates, $G$-band magnitude, parallax, and proper motion. According to the spectral types obtained from {\tt SIMBAD}, about two-third of the stars in the sample belong to the B$-$spectral type.

\begin{table*}
\begin{tabular}{llcccccccc}
\hline
\hline
Nr. & Source & Source Id & RA & Dec. & $G^a$ & Parallax & $f$ & pmra$^b$ & pmdec$^b$ \\
  &  & (\textit{Gaia} DR2)  & $\alpha$ & $\delta$ & Gmag & $\varpi$ & $\sigma_{\varpi}/\varpi$ & $\mu_{\alpha}$\,cos$\delta$ & $\mu_{\delta}$  \\
  &  &   & (deg) & (deg) &  & (mas) &  & (mas y$^{-1}$) & (mas y$^{-1}$)  \\
\hline
1 & HD 227415 & 2058706399118040960 & 300.970271 & 35.566075 & 9.39 & 0.57$\pm$0.03 & 0.05 & -3.85$\pm$0.04 & -6.53$\pm$0.05 \\
2 & HD 227943 & 2058738972149344896 & 302.266859 & 34.729150 & 9.82 & 0.65$\pm$0.04 & 0.06 & -0.73$\pm$0.06 & -2.66$\pm$0.07 \\
3 & HD 228022 & 2058830510807237248 & 302.459886 & 35.429908 & 10.10 & 0.59$\pm$0.04 & 0.07 & -3.03$\pm$0.05 & -7.12$\pm$0.06 \\
4 & HD 228041 & 2058842807280883584 & 302.493409 & 35.496038 & 8.90 & 0.48$\pm$0.04 & 0.08 & -1.43$\pm$0.06 & -2.81$\pm$0.05 \\
5 & HD 191495 & 2058845629092327040 & 302.223101 & 35.512864 & 8.35 & 0.59$\pm$0.05 & 0.09 & -3.12$\pm$0.07 & -7.08$\pm$0.09 \\
6 & HD 191567 & 2058844903225261568 & 302.307975 & 35.484652 & 8.64 & 0.63$\pm$0.04 & 0.07 & -4.21$\pm$0.06 & -6.65$\pm$0.06 \\
7 & HD 228104 & 2058950628143309312 & 302.660170 & 35.874437 & 8.96 & 0.51$\pm$0.03 & 0.06 & -1.19$\pm$0.05 & -3.15$\pm$0.05 \\
8 & HD 191917 & 2058952728407702272 & 302.737574 & 35.953139 & 7.71 & 0.44$\pm$0.04 & 0.08 & -3.41$\pm$0.06 & -6.51$\pm$0.06 \\
9 & HD 228282 & 2058988217688303488 & 303.087437 & 36.346923 & 10.28 & 0.60$\pm$0.06 & 0.10 & -3.30$\pm$0.11 & -5.81$\pm$0.08 \\
10 & HD 190967 & 2059011376158159488 & 301.541434 & 35.385976 & 7.92 & 0.49$\pm$0.03 & 0.07 & -3.44$\pm$0.05 & -6.43$\pm$0.05 \\
11 & HD 227722 & 2059003168499159168 & 301.724294 & 35.306063 & 9.53 & 0.59$\pm$0.03 & 0.05 & -3.52$\pm$0.05 & -6.84$\pm$0.05 \\
12 & BD+35  3976 & 2059033370710817152 & 301.986419 & 35.661458 & 9.87 & 0.58$\pm$0.03 & 0.05 & -2.97$\pm$0.04 & -6.19$\pm$0.05 \\
13 & HD 190919 & 2059072197216667776 & 301.483987 & 35.672053 & 7.19 & 0.43$\pm$0.03 & 0.07 & -3.46$\pm$0.05 & -7.00$\pm$0.05 \\
14 & HD 227634 & 2059075873709364864 & 301.505654 & 35.765502 & 7.85 & 0.54$\pm$0.03 & 0.06 & -3.15$\pm$0.05 & -6.56$\pm$0.05 \\
15 & LS  II +35   32 & 2059076148587293696 & 301.485492 & 35.789922 & 9.66 & 0.59$\pm$0.04 & 0.06 & -3.09$\pm$0.05 & -6.45$\pm$0.06 \\
16 & BD+35  3955 & 2059076251666505728 & 301.494613 & 35.797162 & 7.28 & 0.53$\pm$0.03 & 0.06 & -3.14$\pm$0.05 & -6.29$\pm$0.05 \\
17 & HD 227586 & 2059073159271061632 & 301.401499 & 35.623165 & 8.76 & 0.38$\pm$0.04 & 0.11 & -3.14$\pm$0.07 & -6.37$\pm$0.07 \\
18 & HD 227621 & 2059075255233453824 & 301.469362 & 35.706788 & 10.24 & 0.54$\pm$0.03 & 0.05 & -3.02$\pm$0.04 & -6.33$\pm$0.04 \\
19 & BD+35  3956 & 2059075839349023104 & 301.499889 & 35.762312 & 8.82 & 0.54$\pm$0.03 & 0.06 & -2.93$\pm$0.05 & -6.37$\pm$0.06 \\
20 & HD 190864 & 2059070135632404992 & 301.415827 & 35.607747 & 7.68 & 0.49$\pm$0.03 & 0.06 & -3.07$\pm$0.05 & -6.49$\pm$0.05 \\
21 & HD 227696 & 2059095424401407232 & 301.645676 & 35.740602 & 8.21 & 0.66$\pm$0.04 & 0.06 & -3.21$\pm$0.06 & -6.42$\pm$0.06 \\
22 & HD 227680 & 2059129303105142144 & 301.607110 & 36.329076 & 9.59 & 0.39$\pm$0.03 & 0.08 & -2.75$\pm$0.05 & -5.71$\pm$0.05 \\
23 & HD 227611 & 2059112879149465600 & 301.438097 & 35.900803 & 8.55 & 0.48$\pm$0.03 & 0.06 & -3.15$\pm$0.05 & -6.54$\pm$0.06 \\
24 & HD 191612 & 2059130368252069888 & 302.369194 & 35.733666 & 7.70 & 0.47$\pm$0.04 & 0.08 & -3.59$\pm$0.05 & -5.76$\pm$0.06 \\
25 & HD 227883 & 2059154007751836672 & 302.116375 & 36.093952 & 10.02 & 0.46$\pm$0.03 & 0.06 & -2.98$\pm$0.05 & -5.92$\pm$0.04 \\
26 & HD 227960 & 2059150640497117952 & 302.299000 & 36.048514 & 9.38 & 0.47$\pm$0.03 & 0.07 & -2.78$\pm$0.05 & -5.99$\pm$0.05 \\
27 & HD 191611 & 2059236196250413696 & 302.358642 & 36.488755 & 8.47 & 0.37$\pm$0.04 & 0.10 & -2.57$\pm$0.05 & -4.98$\pm$0.06 \\
28 & HD 227757 & 2059219875373718016 & 301.803210 & 36.359271 & 9.16 & 0.54$\pm$0.03 & 0.05 & 0.37$\pm$0.04 & -3.63$\pm$0.04 \\
29 & HD 191139 & 2059223002094535168 & 301.740030 & 36.396559 & 7.90 & 0.46$\pm$0.03 & 0.07 & -3.67$\pm$0.05 & -6.50$\pm$0.06 \\
30 & BD+36  3882 & 2059225819608656768 & 301.854803 & 36.554469 & 9.80 & 0.54$\pm$0.02 & 0.05 & -3.02$\pm$0.04 & -5.77$\pm$0.04 \\
31 & HD 226868 & 2059383668236814720 & 299.590295 & 35.201580 & 8.52 & 0.42$\pm$0.03 & 0.08 & -3.88$\pm$0.05 & -6.17$\pm$0.05 \\
32 & HD 227245 & 2059455613211582080 & 300.590475 & 35.674923 & 9.47 & 0.63$\pm$0.03 & 0.04 & -3.96$\pm$0.04 & -6.42$\pm$0.05 \\
33 & HD 227018 & 2059434898612310656 & 299.954565 & 35.309285 & 8.86 & 0.47$\pm$0.03 & 0.07 & -5.59$\pm$0.05 & -7.99$\pm$0.05 \\
34 & LS  II +35   21 & 2059458778631320064 & 300.869369 & 35.692847 & 10.30 & 0.64$\pm$0.02 & 0.04 & -3.54$\pm$0.04 & -6.34$\pm$0.04 \\
35 & HD 227132 & 2059524543174097664 & 300.311790 & 35.956635 & 10.02 & 0.40$\pm$0.02 & 0.06 & -2.47$\pm$0.04 & -5.29$\pm$0.04 \\
36 & HD 226951 & 2059706203121650432 & 299.798166 & 36.115305 & 9.08 & 0.50$\pm$0.03 & 0.06 & -4.22$\pm$0.05 & -6.11$\pm$0.05 \\
37 & HD 227070 & 2059744720391006720 & 300.117591 & 36.532764 & 10.00 & 0.40$\pm$0.03 & 0.08 & -0.83$\pm$0.05 & -6.79$\pm$0.06 \\
38 & TYC 2682-2524-1 & 2059878001807789952 & 301.482346 & 36.274201 & 10.10 & 0.43$\pm$0.03 & 0.07 & -3.07$\pm$0.05 & -5.95$\pm$0.05 \\
39 & BD+35  3929 & 2059854121788578432 & 300.864567 & 36.034972 & 9.39 & 0.54$\pm$0.03 & 0.06 & -2.88$\pm$0.05 & -5.87$\pm$0.05 \\
40 & HD 227460 & 2059863536357618816 & 301.067545 & 36.265368 & 9.46 & 0.56$\pm$0.03 & 0.06 & -3.03$\pm$0.05 & -6.13$\pm$0.05 \\
41 & HD 227607 & 2059885801468671744 & 301.439600 & 36.519983 & 9.88 & 0.52$\pm$0.03 & 0.05 & -4.36$\pm$0.04 & -6.26$\pm$0.05 \\
\hline
\hline
\end{tabular}
\caption{\textit{Gaia} data for the 41 stars in our sample. (a) $G$ denotes the G-band magnitudes of the stars obtained from \textit{Gaia} DR2. (b) pmra and pmdec represent the proper motions of the stars in RA and Dec.}
\label{tab1}
\end{table*}

\begin{table*}
\begin{tabular}{cccccccc}
\hline
\hline
Nr. & Source & Spectral type & $r\sub{exp}$ & \vpec{} & \vrel{} & $v\sub{r}$\\

&  &  & (kpc) & (km s$^{-1}$) & (km s$^{-1}$) &  (km s$^{-1}$)\\
\hline
1 & HD 227415 & B3 & 1.76$\pm$0.08 & 28.3$\pm$7.8 & 21.4$\pm$10.0 & -25.0$\pm$10.0 \\
2 & HD 227943 & B8 & 1.54$\pm$0.09 & -- & -- & -- \\
3 & HD 228022 & B3:III: & 1.70$\pm$0.12 & -- & -- & -- \\
4 & HD 228041 & B0.5V:e & 2.09$\pm$0.17 & -- & -- & -- \\
5 & HD 191495 & B0IV-V(n) & 1.70$\pm$0.15 & 23.2$\pm$4.9 & 17.0$\pm$5.5 & 5.0$\pm$3.7 \\
6 & HD 191567 & B1V & 1.59$\pm$0.11 & 30.5$\pm$5.0 & 24.6$\pm$5.1 & -29.0$\pm$1.8 \\
7 & HD 228104 & B1:IV:pe & 1.98$\pm$0.12 & -- & -- & -- \\
8 & HD 191917 & B1III & 2.26$\pm$0.19 & 22.7$\pm$5.3 & 13.6$\pm$5.3 & -18.0$\pm$3.7 \\
9 & HD 228282 & B8 & 1.69$\pm$0.17 & -- & -- & -- \\
10 & HD 190967 & O9.5V+B1Ib & 2.03$\pm$0.14 & 20.8$\pm$4.8 & 8.8$\pm$4.2 & -16.1$\pm$1.8 \\
11 & HD 227722 & B1III & 1.70$\pm$0.08 & 21.8$\pm$4.9 & 13.0$\pm$7.0 & -1.0$\pm$7.4 \\
12 & BD+35 3976 & OB- & 1.74$\pm$0.09 & -- & -- & -- \\
13 & HD 190919 & B0.7Ib & 2.32$\pm$0.15 & 26.2$\pm$5.2 & 11.3$\pm$4.4 & -15.0$\pm$3.7 \\
14 & HD 227634 & B0.2II & 1.85$\pm$0.11 & 17.5$\pm$4.5 & 7.7$\pm$3.6 & -10.0$\pm$3.7 \\
15 & LS II +35 32 & B1V & 1.71$\pm$0.11 & -- & -- & -- \\
16 & BD+35 3955 & B0.7Iab & 1.88$\pm$0.11 & 15.3$\pm$4.3 & 9.1$\pm$4.3 & -6.0$\pm$1.8 \\
17 & HD 227586 & B0.5IVp & 2.63$\pm$0.32 & 19.6$\pm$5.3 & 12.0$\pm$5.8 & -6.0$\pm$7.4 \\
18 & HD 227621 & B1.5IV & 1.87$\pm$0.09 & -- & -- & -- \\
19 & BD+35 3956 & B0.5Vne & 1.85$\pm$0.11 & 19.7$\pm$6.5 & 13.0$\pm$6.3 & -19.0$\pm$7.4 \\
20 & HD 190864 & O6.5III(f) & 2.05$\pm$0.13 & 17.6$\pm$4.6 & 11.5$\pm$4.9 & -2.8$\pm$3.6 \\
21 & HD 227696 & B1III & 1.52$\pm$0.09 & 17.0$\pm$4.7 & 8.8$\pm$4.3 & -15.0$\pm$3.7 \\
22 & HD 227680 & B3II & 2.56$\pm$0.21 & -- & -- & -- \\
23 & HD 227611 & B1:III/Ve & 2.08$\pm$0.13 & 25.9$\pm$8.2 & 15.1$\pm$8.3 & -24.0$\pm$10.0 \\
24 & HD 191612 & O8fpe & 2.14$\pm$0.17 & 26.3$\pm$5.4 & 17.3$\pm$5.2 & -27.6$\pm$3.3 \\
25 & HD 227883 & B & 2.19$\pm$0.14 & -- & -- & -- \\
26 & HD 227960 & O8.5 & 2.14$\pm$0.15 & 12.3$\pm$4.4 & 11.2$\pm$4.3 & -8.0$\pm$3.7 \\
27 & HD 191611 & B0.5III & 2.71$\pm$0.27 & 13.4$\pm$5.5 & 26.6$\pm$5.8 & 4.0$\pm$3.7 \\
28 & HD 227757 & O9.5V & 1.87$\pm$0.09 & 28.3$\pm$4.9 & 43.8$\pm$4.7 & -16.0$\pm$7.0 \\
29 & HD 191139 & B0.5III & 2.19$\pm$0.16 & 22.2$\pm$4.9 & 8.2$\pm$3.9 & -13.0$\pm$3.7 \\
30 & BD+36 3882 & B1III: & 1.85$\pm$0.08 & -- & -- & -- \\
31 & HD 226868 & O9.7Iabpvar & 2.38$\pm$0.19 & 22.2$\pm$4.3 & 11.4$\pm$3.0 & -5.1$\pm$0.5 \\
32 & HD 227245 & O7 & 1.60$\pm$0.07 & 21.1$\pm$4.2 & 9.7$\pm$3.2 & -13.0$\pm$3.7 \\
33 & HD 227018 & O6.5III & 2.12$\pm$0.15 & 52.2$\pm$5.2 & 42.4$\pm$5.6 & 20.0$\pm$4.4 \\
34 & LS II +35 21 & B & 1.55$\pm$0.06 & -- & -- & -- \\
35 & HD 227132 & B2III & 2.48$\pm$0.15 & -- & -- & -- \\
36 & HD 226951 & B0.5III & 1.99$\pm$0.13 & 23.4$\pm$4.0 & 11.7$\pm$2.4 & -9.0$\pm$2.9 \\
37 & HD 227070 & B2 & 2.54$\pm$0.20 & -- & -- & -- \\
38 & TYC 2682-2524-1 & B5? & 2.36$\pm$0.16 & -- & -- & -- \\
39 & BD+35 3929 & B1III & 1.85$\pm$0.11 & -- & -- & -- \\
40 & HD 227460 & B0.5:V & 1.78$\pm$0.10 & 14.2$\pm$4.4 & 8.3$\pm$3.9 & -10.0$\pm$3.7 \\
41 & HD 227607 & B1:Ib: & 1.92$\pm$0.09 & 27.6$\pm$5.2 & 19.8$\pm$6.7 & 5.0$\pm$7.4 \\
\hline
\end{tabular}
    \caption{Spectral types of the stars mentioned in the third column are obtained from {\tt SIMBAD}. The distances, $r\sub{exp}$, in the fourth column are the heliocentric distances to the stars calculated using exponential prior. \vpec{} mentioned in the fifth column are the peculiar velocities of stars after removing the components of Solar motion and Galactic rotation. \vrel{} are the relative velocities of individual stars with respect to the association. Radial velocities \vr{} given in the last column are obtained from \citet{kharchenko07}.}
\label{table:tab2}
\end{table*}


\section{Distance measurements}
\label{distancefinding}

Heliocentric distances can be estimated by inverting the parallaxes provided in Table~\ref{tab1} for individual stars ($r\sub{inv}$ = 1/ $\varpi $). \cite{bailer-jones15} and \citet{astraat16} emphasized that parallax inversion may not serve as a good estimator to find the distance and its uncertainty interval if the fractional uncertainty in parallax measurement is large ($\gtrsim$ 0.2). It is observed that the fractional uncertainty is less than 0.2 for most of our sample (Table~\ref{tab1}).

We estimated the distances to individual stars with a Bayesian prior that considers an exponential decrease of space density as in \citet{bailer-jones15} with a length scale of 2 kpc corresponding to the approximate mean distance to \ob{} and compared with the distances obtained from parallax inversion. The two distance estimates are consistent with each other within 2$\sigma$ level,
which was expected considering the smaller fractional uncertainties in the parallax measurements. We compared the distance estimates from exponential prior with a length scale of 1.35 kpc used in \cite{astraat16}, which were found to be consistent with those obtained with a length scale of 2 kpc. We also searched for our sources in the catalogue provided by \citet{anders19}, where the authors estimate accurate distances to a good number of sources using parallaxes from \textit{Gaia} DR2 and photometric catalogues of Pan-STARRS1, 2MASS, and AllWISE. The distances are found to be consistent with those obtained from \citet{anders19} within uncertainties. Distances are provided in Table~\ref{table:tab2}.

\begin{figure}
	\centering
	\includegraphics[height=7cm, width=7 cm]{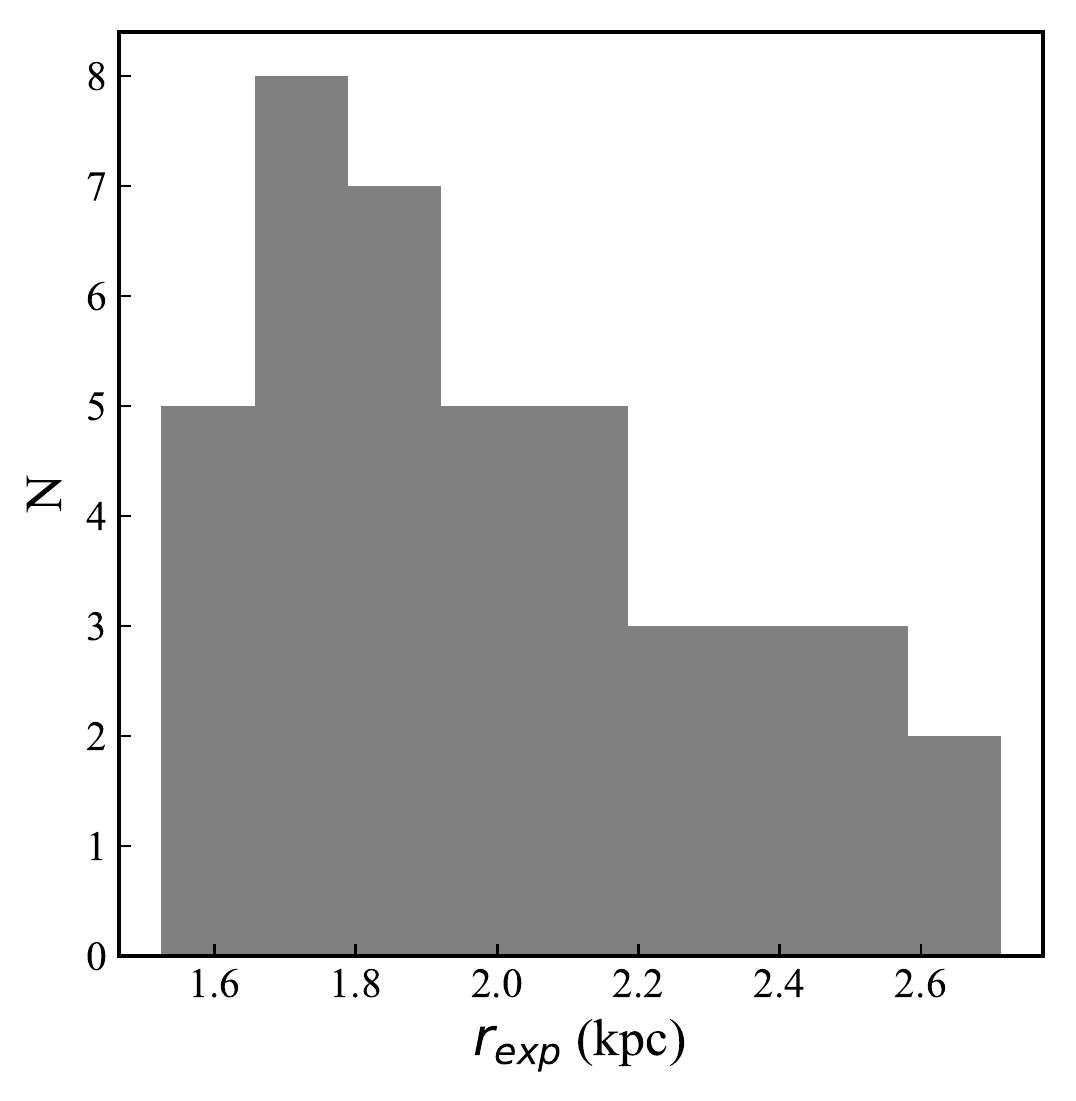}
    
	\caption{Histogram of heliocentric distance estimates, $r\sub{exp}$, obtained from exponential prior.}
	\label{dist_histo}
\end{figure}

A histogram of the distances ($r\sub{exp}$) is shown in Fig~\ref{dist_histo}; the majority of stars are located at a distance of about 2 kpc, suggesting a higher density of OB-type stars in a region spanning 0.31 kpc equal to the standard deviation of $r\sub{exp}$ estimates. In order to compare the observed width of the parallax distribution and the expected width entirely due to parallax uncertainties, we generated 500 realizations of parallax distributions for 41 sources by drawing random values from a normal distribution defined by a mean <$\varpi$> = 0.5 mas with the respective $\sigma_\varpi$. The observed parallax distribution is found to be significantly wider than the ones obtained from simulations, and suggests that the observed scatter of sources along the line of sight is real and is not entirely driven by parallax uncertainties.

Histograms were once again simulated with parallax uncertainties corresponding to the size of the association (assumed to be 100 pc for a distance of 2 kpc) and the intrinsic variance defined as $\sigma\sub{int}^2$ =  $\sigma\sub{obs}^2 - \sigma\sub{sim}^2$ was estimated, where $\sigma\sub{obs}$ and $\sigma\sub{sim}$ are, respectively, the widths of the observed and simulated histograms. It is found that the parallax uncertainties in the calculation of $\sigma\sub{sim}$ would have to be increased by a factor of 3 to get zero intrinsic variance. This is unlikely to be produced by systematic effects and therefore, is suggestive of either a real extension of the stars or contamination by non-members. However, if we adopt the maximum global systematic uncertainty of 0.1 mas stated by the \gaia{} team \citep{luri18}, the result is consistent with zero intrinsic variance.


The mean and median distance for the sample are found to be, respectively, 1.99 kpc and 1.92 kpc, with a standard deviation of 0.31 kpc. The distance of 1.92$\pm$0.31 kpc for the association \ob{} is in agreement with the findings of \citet{dambis01}, \citet{melnik01}, \citet{mirabel-rodriguez-2003}, \citet{melnik95}, and \citet{melnik17}. The distance to \cygx{} is found to be 2.38$\pm$0.19 kpc \citep{gandhi19}, consistent with the distance to the association. The nearest stars are at a distance of about $\sim$1.5 kpc, as constrained by the minimum distance in the archive query.

\begin{figure}
	\centering
    \includegraphics[height=7cm, width=7 cm]{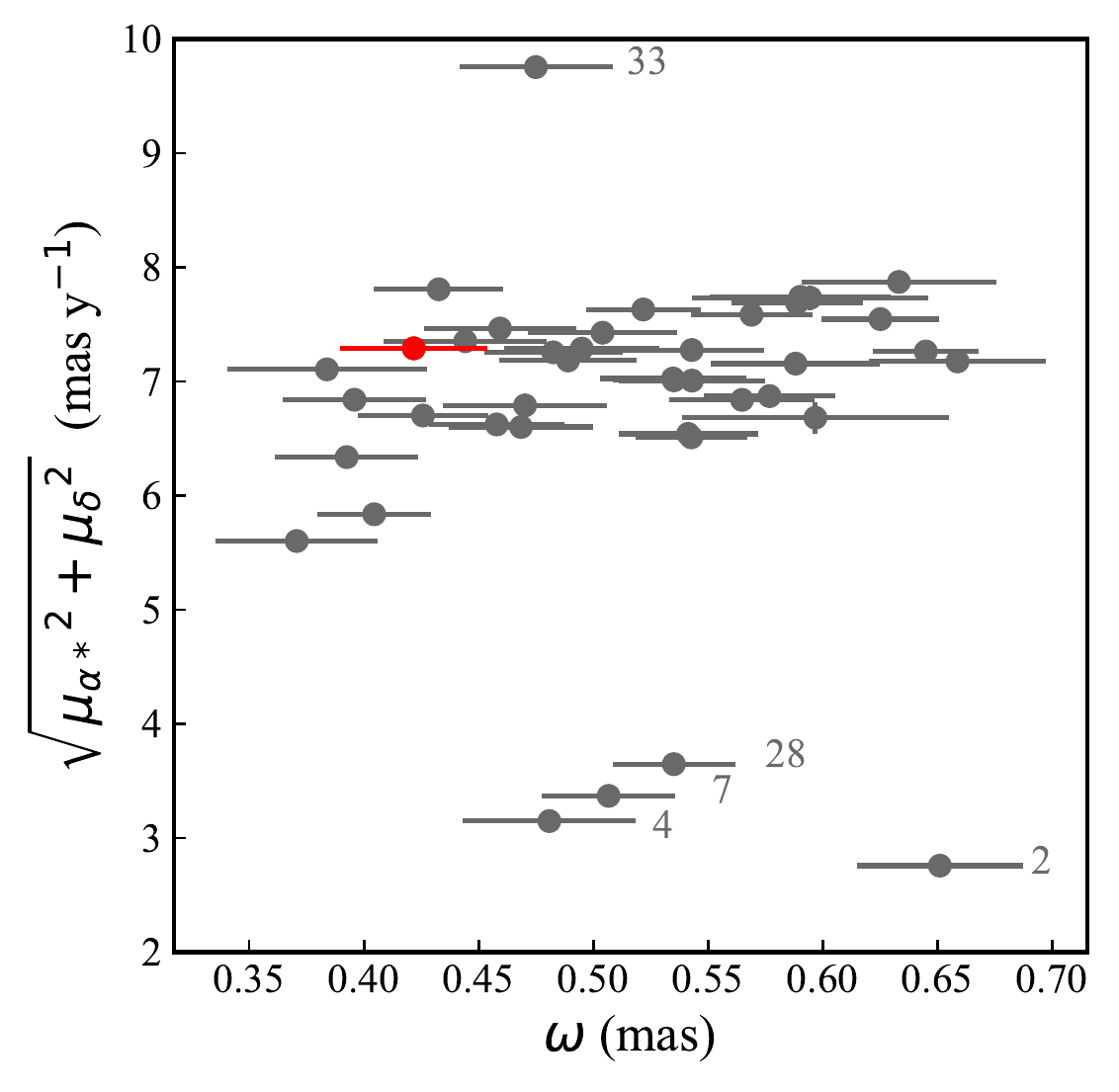}  
	\caption{Parallax as a function of proper motion for the stars in our sample. Cyg X-1 is shown in red. The serial numbers for some of the stars from Table~(\ref{tab1}) are shown in the figure.}
	\label{pm}
\end{figure}

\section{Proper Motions}
\label{secpm}
\textit{Gaia} DR2 provides proper motion measurements for a large number of stars in the region covered by \ob{}. Proper motions are defined as $\mu~= \sqrt{\mu_{\alpha_{\ast}}^{2} + \mu_{\delta}^{2}} $, where $\mu_{\alpha_{\ast}}$ ( = $\mu_{\alpha}$ $\cos\delta$) and $\mu_{\delta}$ are proper motions along RA ($\alpha$) and Dec ($\delta$), respectively.  The proper motions are shown as a function of parallax in Fig~\ref{pm}. Since nearby stars usually exhibit higher proper motions, we expect to find a correlation between parallax and proper motion, as observed in the figure. However, a significantly higher proper motion for HD 227018 [{\tt 33}] and lower values for four of the stars viz. HD~227943 [{\tt 2}], HD~228041 [{\tt 4}], HD~228104 [{\tt 7}] and HD~227757 [{\tt 28}] marked in the figure are evident outliers. The number mentioned in parenthesis indicates the index of the star according to the serial number in Tables~\ref{tab1} and \ref{table:tab2}. \cygx{} does not exhibit any peculiarity in its proper motion, when compared with the other stars. We investigated the presence of kinematic substructures in the sample based on their proper motion using Shapiro-Wilk test. The test suggests that the distributions of $\mu_{\alpha_{\ast}}$ and $\mu_{\delta}$ do not follow single normal distributions, and hence indicate the presence of kinetic subgroups. The distributions are consistent with single normal distributions when stars {\tt 2}, {\tt 4}, {\tt 7}, {\tt 28}, {\tt 33}, and {\tt 37} are removed from the test.

\begin{figure}
	\centering
	\includegraphics[height=7.5cm, width=7.7 cm]{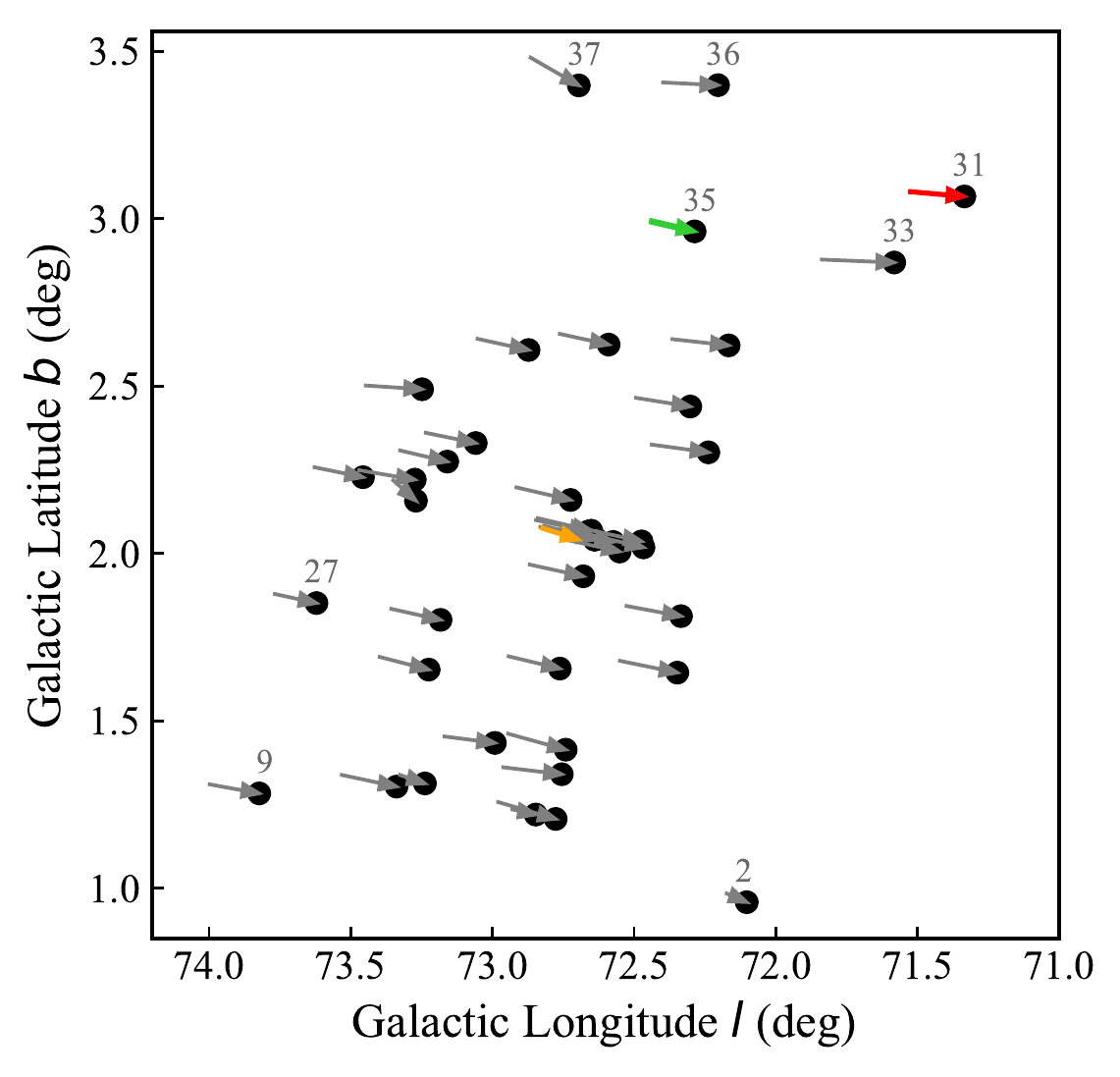}
\caption{Stars are shown in the Galactic $l-b$ plane. The grey arrows represent their proper motions over the past 0.1 Myr. Serial numbers for some of the stars from Table~(\ref{tab1}) are marked in the figure. The red and green arrows represent $\mu$ for \cygx{}~[{\tt 31}] and HD 227018 [{\tt 33}], respectively. The proper motion of the center of the association is shown via the yellow arrow.}
	\label{propermotion}
\end{figure}

Fig~\ref{propermotion} shows the proper motions of the stars in the $l-b$ plane. The starting positions of the arrows correspond to their locations extrapolated back in time by 0.1 Myr, with the arrowheads shown at their present positions. The length of each arrow is proportional to the proper motion of the respective star. The serial numbers of some of the stars from Tables~\ref{tab1} and \ref{table:tab2} are shown in the figure. Not all stars are marked for the sake of clarity. Although the proper motions of most of the stars point in a similar direction and have similar magnitudes, some anomalies are observed. For example, the smaller proper motion of HD 227943 [{\tt 2}] is noticeable in the figure. 

We also calculate the proper motion of the association by determining the median $l-b$ coordinates of the stars and extrapolating into the past using the median proper motion of the stars. The proper motion of the association is presented with a yellow arrow in the figure. Note that the yellow arrow is very similar to that of \cygx{}, shown in red, both in terms of direction and magnitude. This result, obtained from \textit{Gaia} DR2 data for 41 stars, is consistent with the findings of \citet{mirabel-rodriguez-2003}, who studied the association using Hipparcos data for 22 stars. 

Galactic rotation and Solar motion significantly contribute to the observed proper motions of stars in the sky. Although the study of the proper motions of any sample provides important clues about the peculiarity in the kinematics of stars, it is important to exclude these two strong contributors to the proper motion in order to find the intrinsic motion of the stars. Therefore, next we calculate the peculiar velocities of the stars in our sample, accounting for the full three-dimensional motions of the stars, including the line of sight, radial velocity.

\section{Peculiar velocities} 
\label{vpecfinding}

The estimation of the peculiar velocity relative to Galactic rotation and Solar motion for any star requires a full astrometric solution, i.e. parallax, proper motion and radial velocity ($v\sub{r}$), as input. \textit{Gaia} has a spectrometer on board that can measure radial velocities, however none of the stars studied in this work have measured DR2 $v\sub{r}$ values. 
Therefore, these measurements were extracted from the catalogue published by \citet{kharchenko07}. The catalogue lists 25 of the stars given in Table~\ref{tab1} for which peculiar velocities (\vpec{}) and subsequent inferences are discussed herein. The $v\sub{r}$ for \cygx{} is taken from \cite{gies08}.
The formalism of \citet{johnson87} is used to obtain the heliocentric space velocities ($U, V, W$). Peculiar velocities were calculated by removing Solar motion and Galactic rotation according to the formalism and constants given in \cite{reid09,reid14} and is defined as the space velocity after removing Galactic rotation and Solar motion, $v\sub{pec}$ = $\sqrt {U\sub{s}^{2} + V\sub{s}^{2} + W\sub{s}^{2} }$, where , \us{}, \vs{} and \ws{} are the velocity components towards the Galactic centre, along the direction of Galactic rotation, and towards the North Galactic pole.

The mean and standard deviation of 10,000 realizations of \vpec{} have been quoted in Table~\ref{table:tab2} by drawing random values of $\varpi$, $\mu$ and $v\sub{r}$, accounting for the covariance between $\varpi$ and $\mu$ quoted in the \textit{Gaia} DR2 astrometric solution. The calculation of \vpec{} also accounts for the uncertainties of the circular rotation speed of the Galaxy at the location of the Sun ($\Theta_0$), Solar motion towards the Galactic centre ($U_\odot$), Solar motion in the direction of Galactic rotation ($V_\odot$), Solar motion towards the north Galactic pole ($W_\odot$) and the distance of the Sun from the Galactic centre ($R_\odot$) from \cite{reid14}.

 \begin{figure}
	\centering
	\includegraphics[height=5.81cm]{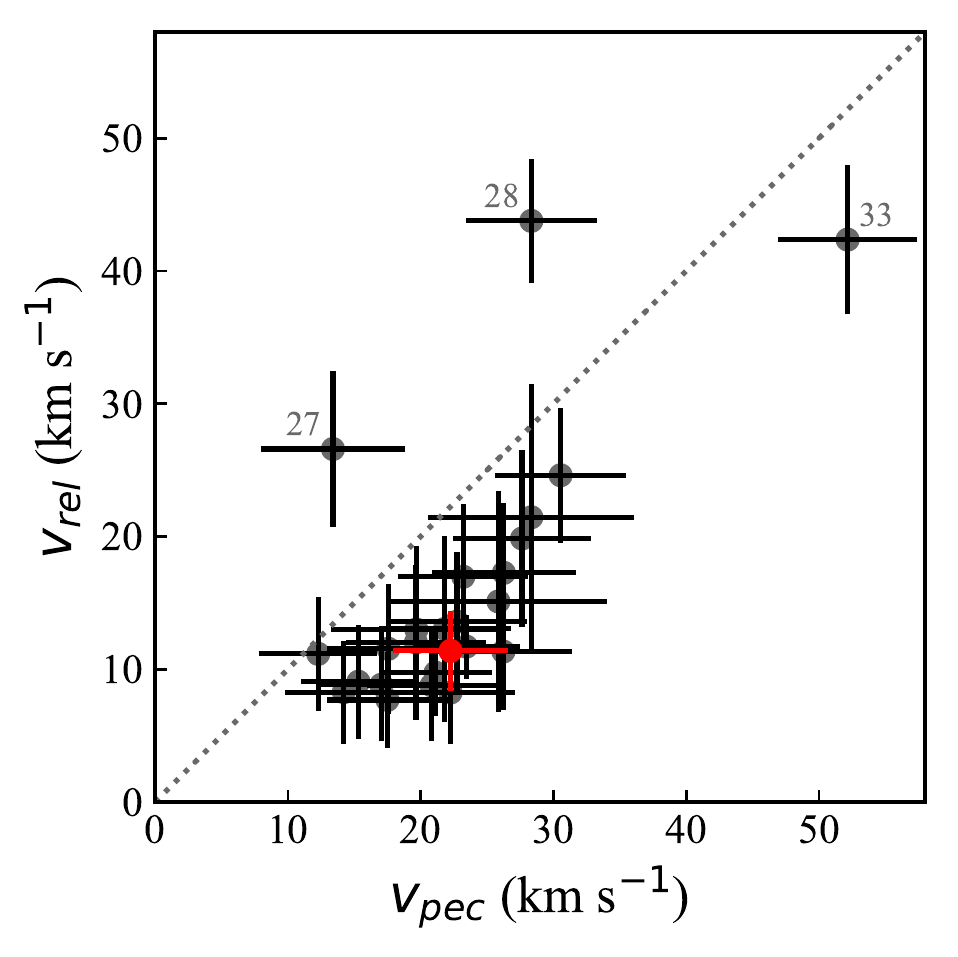}

	\caption{A comparison between the velocities of stars relative to the association and their peculiar velocities is shown. The dotted line corresponds to equality between the two quantities. The serial numbers for some of the stars from Table~(\ref{tab1}) are shown in the figure. \cygx{} is shown in red.}
	\label{vpec}
	\end{figure}

A majority of the stars exhibit \vpec{} of $\sim$ 22 \kms{}, with a median value of 22 \kms{} and a range of 10-30 \kms{}, except HD~227018 [{\tt 33}] that has a peculiar velocity of $\sim$52 \kms{}. The velocity dispersion, measured as the \vpec{} standard deviation is equal to 7.7 \kms{}. This velocity dispersion is small considering the typical \vpec{} uncertainties of individual stars given in Table~\ref{table:tab2} of about 6 \kms{}. This is consistent with \citet{melnik17} and \citet{melnik95}.

The peculiar velocity of Cyg X-1 is found to be 22.2~$\pm$~4.3 \kms{}, consistent with the mean of the association. This result is consistent with the findings of \citet{mirabel-rodriguez-2003} that the two systems have low peculiar velocities \citep[see also][]{gandhi19}. The match also indicates that Cyg X-1 is a member of the common group formed by the stars in the region. This is consistent with the inferences drawn by \citet{mirabel17-review} and  \citet{mirabel-rodriguez-2003} regarding the membership of \cygx{} in \ob{}.

We also calculate the relative three-dimensional velocities \vrel{} of individual stars with respect to the association defined as $v\sub{rel}$ = $\sqrt {(U\sub{s}- U\sub{m} )^{2} + (V\sub{s}- V\sub{m} )^{2} + (W\sub{s}- W\sub{m} )^{2} }$, where $U$\sub{m}, $V$\sub{m}, $W$\sub{m} are the median velocity components of the association. A majority of the stars exhibit \vrel{} of about 15 \kms{}. The mean and standard deviation of 10,000 \vrel{} realizations for each star
are shown in Table~\ref{table:tab2}. Astrometric covariances are taken into account. The mean \vrel{} is 15.9~$\pm$~9.4 \kms{}, where the quoted uncertainty is the standard deviation of the relative velocities of 25 stars. 

The relative velocity of \cygx{} is estimated to be 11.4~$\pm$~3.0 \kms{}, consistent with an estimate of 9$\pm$2 \kms{} by \cite{mirabel-rodriguez-2003}. 
We compare the peculiar velocities of stars with their relative velocities and the results are shown in Fig~\ref{vpec}. 
A positive correlation is observed between the two quantities. Most of the stars show \vrel{} $<$ \vpec{}, however a few exceptions are apparent viz.  HD~191611 [{\tt 27}], and HD~227757 [{\tt 28}].

 \begin{figure}
	\centering
	\includegraphics[height=6.2cm, width=6.81 cm]{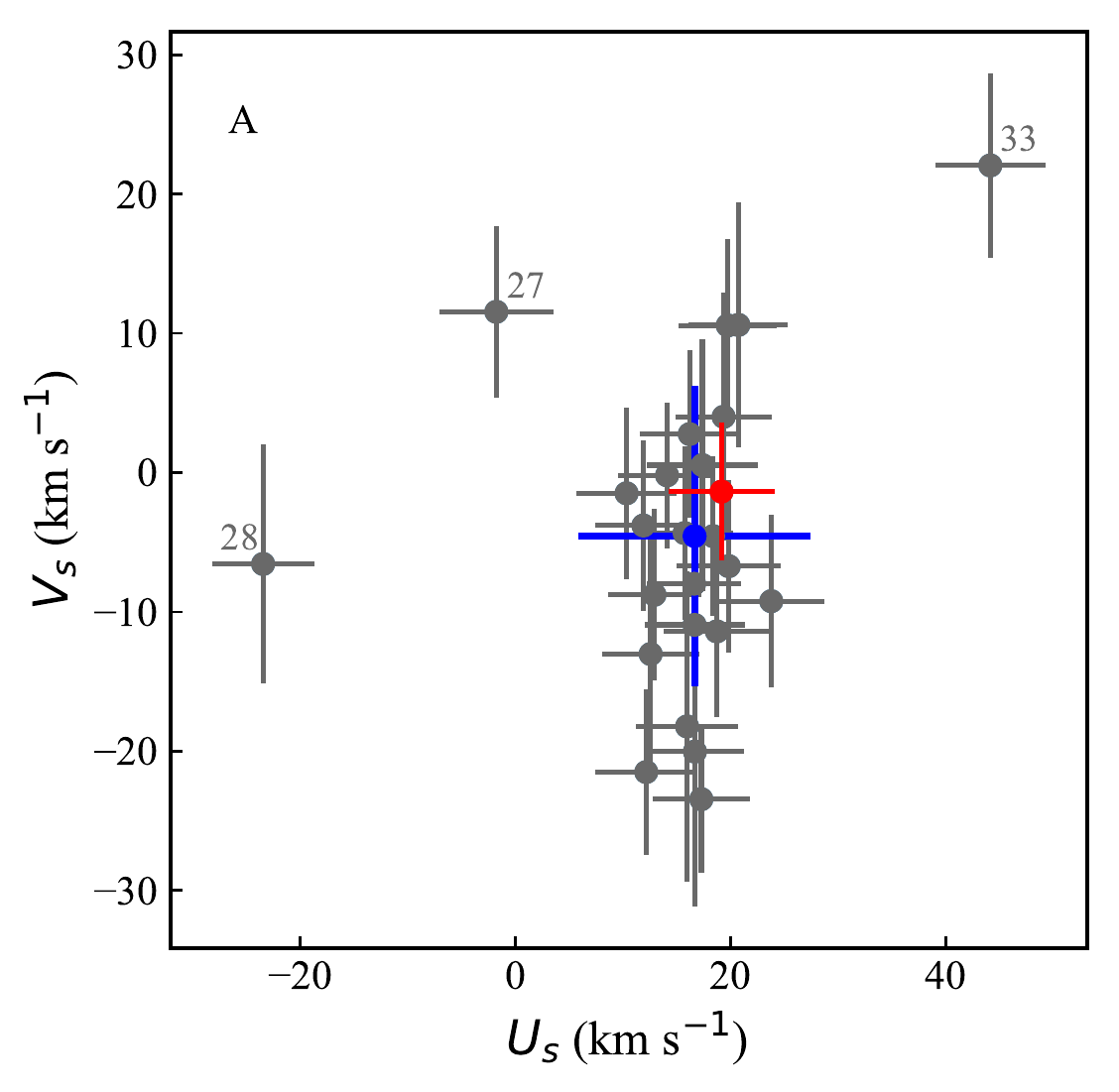}
	\includegraphics[height=6.2cm, width=6.81 cm]{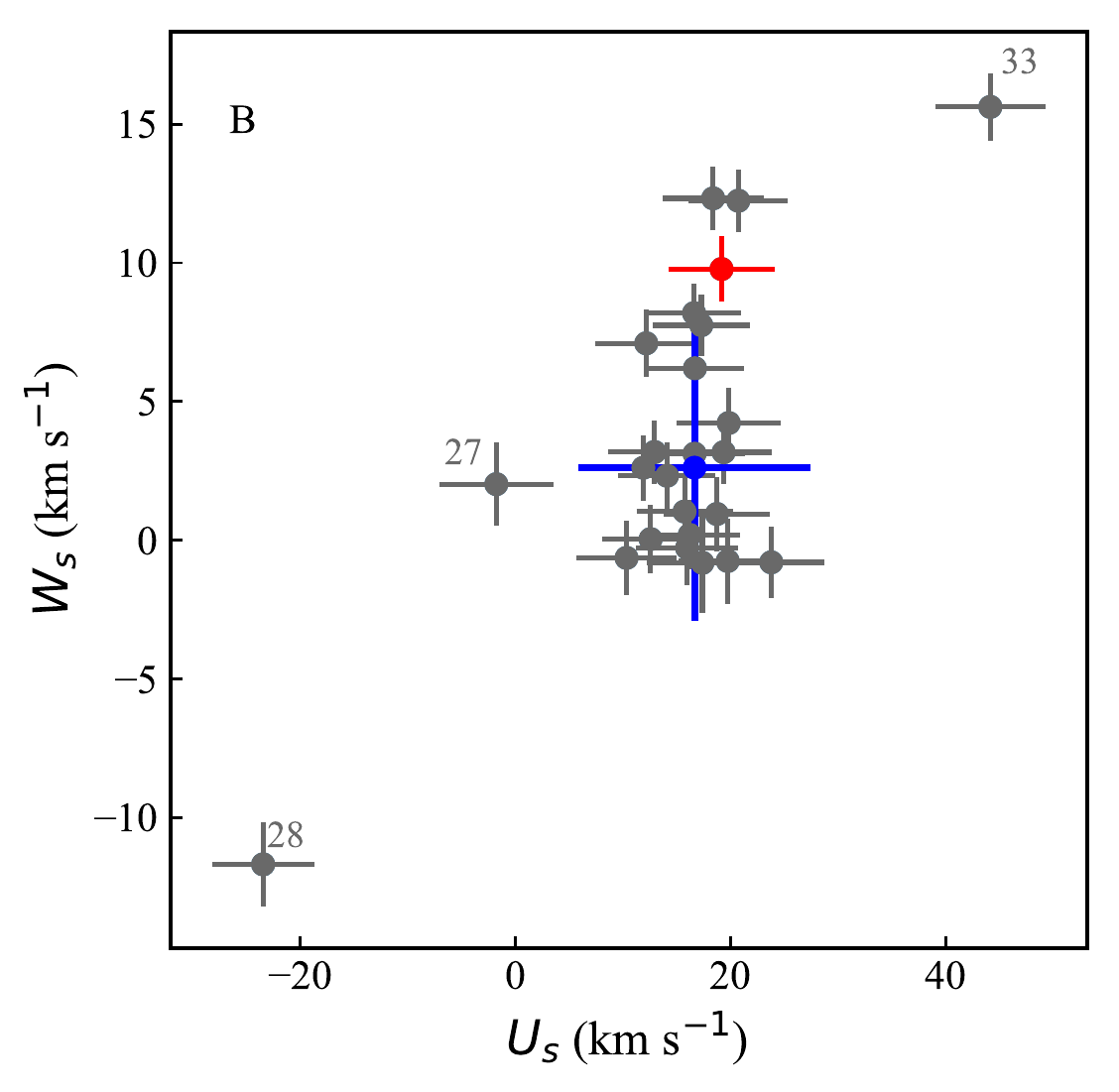}
	\includegraphics[height=6.2cm, width=6.81 cm]{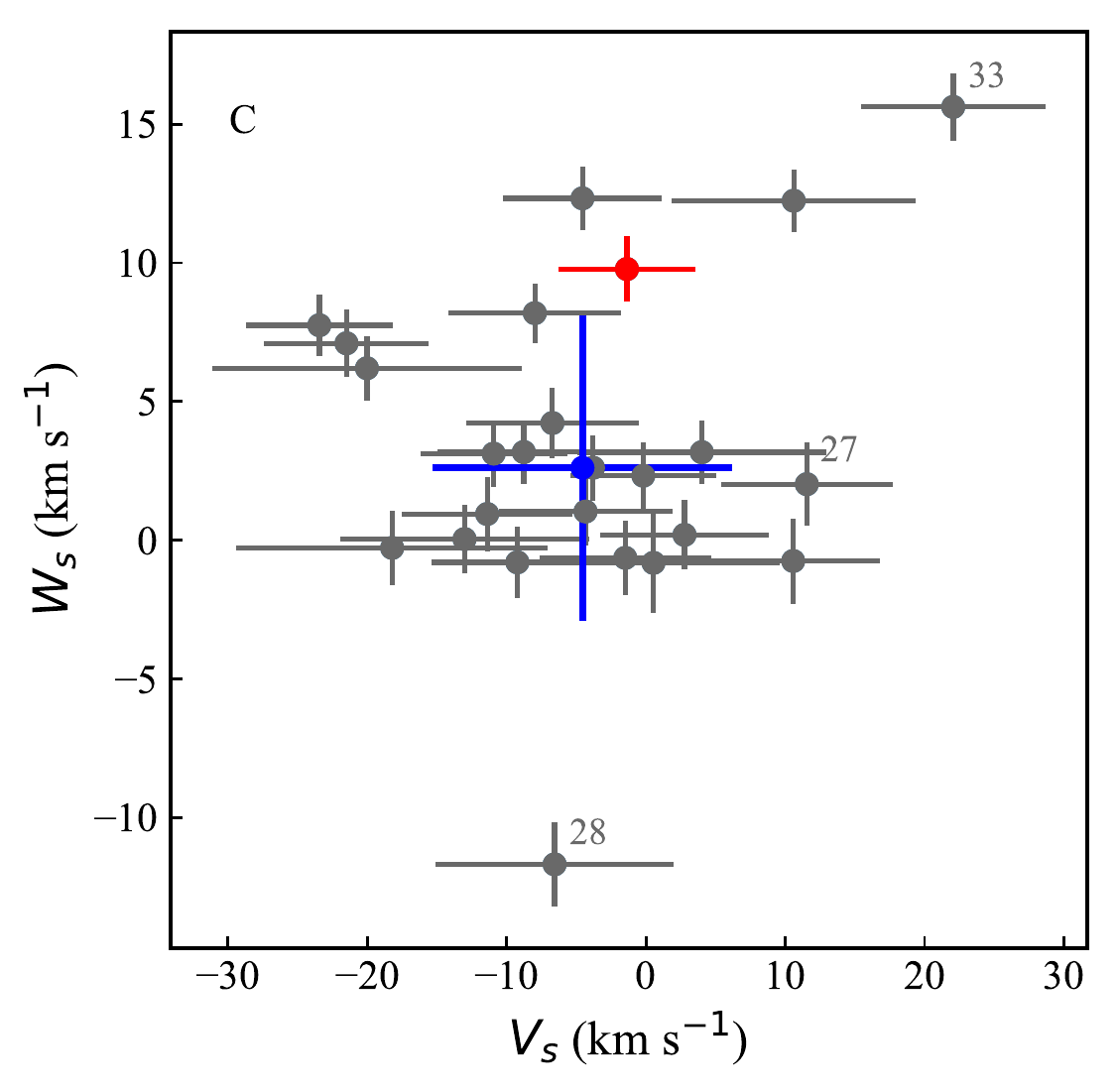}
	
\caption{The figure shows components of space velocities \us{}, \vs{} and \ws{}. \cygx{} is shown in red. Median values of the components and their standard deviation are shown in blue. Serial numbers marked for some of the stars are from Table~\ref{tab1}.}
	\label{usvs}
\end{figure}

\subsection{Peculiar velocities (\us{}, \vs{} and \ws{}) }

We study the components of \vpec{} viz. \us{}, \vs{} and \ws{} for our 25 stars, shown in Fig~\ref{usvs}. The median values of the components that can be considered as the velocity components of the association are found to be ($U$\sub{m}, $V$\sub{m}, $W$\sub{m}) = (16.6\er{}10.8, $-$4.6\er{}10.8, 2.6\er{}5.5) \kms{}, where the quoted uncertainties are the standard deviation for the sample.
\vpec{} is dominated by the \us{} component along the direction to the Galactic centre, which has a median value of 16.6 \er{} 10.8 \kms{} for the sample. The first panel in Fig.~\ref{usvs} shows that the stars move as a coherent group in this direction, excluding three stars marked in the figure with their serial numbers. The standard deviation in \us{} reduces to 3.2 \kms{}, when the three stars are not considered. 
A majority of the stars are observed moving towards the North Galactic pole with a median velocity of 2.6 \er{} 5.5 \kms{} for the sample. The two stars HD~191611 [{\tt 27}] and HD~227757 [{\tt 28}], exhibiting higher relative velocities in Fig~(\ref{vpec}), have a velocity component \us{} that deviates from the rest of the stars. The velocity components for the runaway star HD~227018 [{\tt 33}] are the highest in all three directions. The velocity components for \cygx{} are found to be (\us{}, \vs{}, \ws{}) = (19.1\er{}4.9, $-$1.4\er{}4.9, 9.8\er{}1.2) \kms{} and it appears to be moving with the coherent structure formed by other stars in the association.

\section{Evolution of the association}

A publicly available tool {\tt galpy} \citep{bovy15} allows the calculation of stellar orbits under a variety of Galactic potentials, and extrapolation of orbits in time. We make use of this facility to study the distance of individual stars with respect to the center of the association ($r\sub{rel}$) at present, and also to study the projected RA and Dec positions after 10 Myr. The tool requires a six-parameter input as initial conditions to integrate the orbits viz. $\alpha$, $\delta$, $r$, $\mu_{\alpha}$ cos$\delta$, $\mu_{\delta}$, and $v\sub{r}$. The calculation accounts for the Galactic potential, but does not consider the gravitational potential of the association or surrounding interstellar medium as these are expected to be unimportant (see below).
The orbits of individual stars are integrated 1000 times by drawing random values of $\varpi$, $\mu$ and \vr{}. The covariance between $\varpi$ and $\mu$ has been taken into account.  
Most of the stars are found to be within $\sim$ 500 pc from the centre of the association at present, though the uncertainties on $r\sub{rel}$ are large. The results of our error analysis reveal that the parallax is the most significant contributor to the uncertainties in $r\sub{rel}$ estimates. 

\begin{figure}
	\centering
	\includegraphics[height=6.2cm, width=6.8 cm]{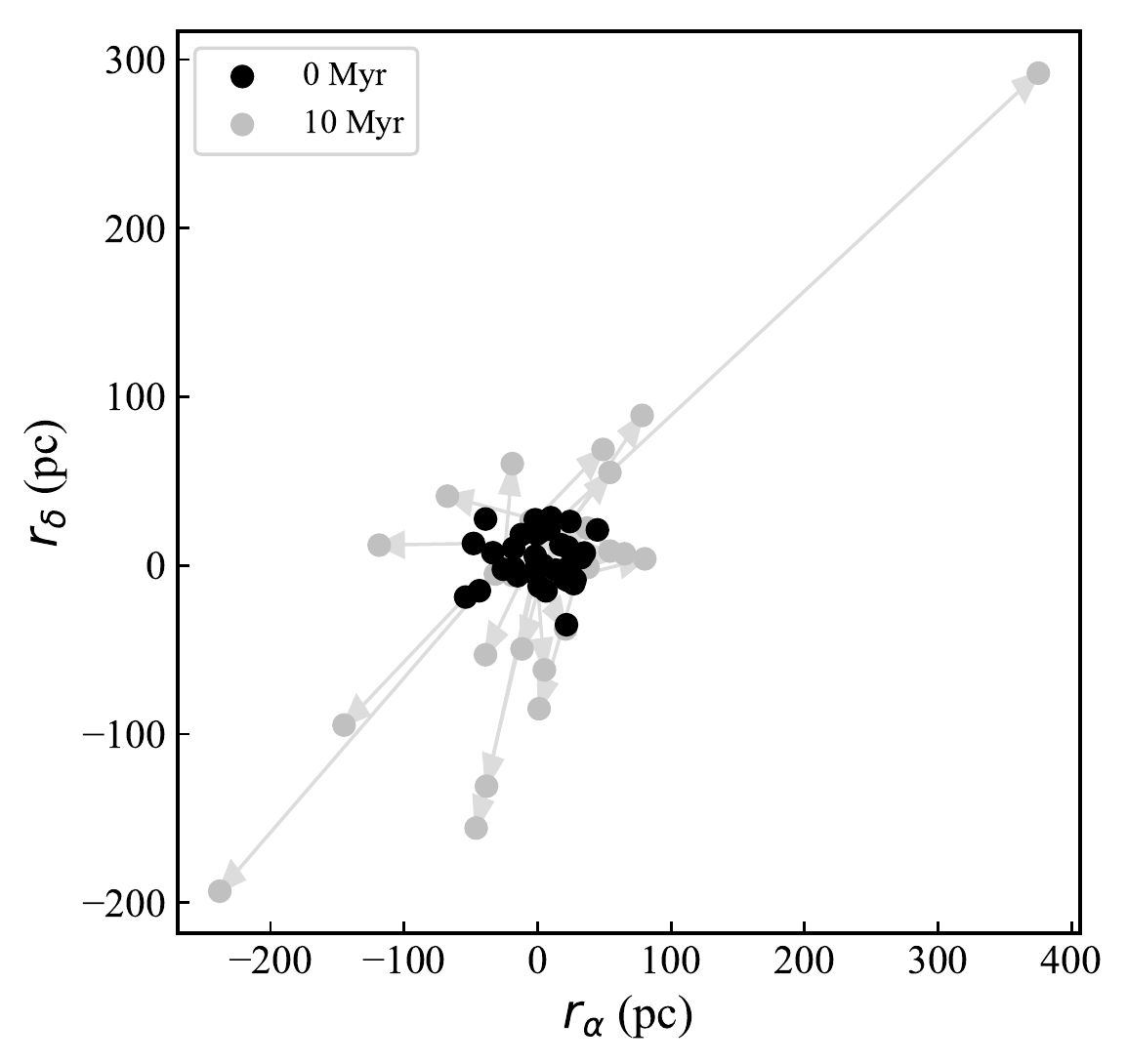}

    \caption{Extrapolated expansion of the association in the sky plane when only the Galactic potential is considered. $r_\alpha$ and $r_\delta$ are the distances from the centre of the association along the RA and Dec directions.}
	\label{skyexpansion}
\end{figure}

We study the current and projected RA and Dec positions over the next 10 Myr on the sky plane on parsec scales using {\tt galpy} assuming all stars lie at the median association distance. The current positions for 41 stars and the projected positions for 25 of the stars are shown in Fig~\ref{skyexpansion} in black and grey, respectively. $r_\alpha$ and $r_\delta$ are the distances from the centre of the association along RA and Dec. The light grey arrows connect the corresponding stars at T=0 and T=10 Myr. The figure suggests an expansion of the association in the plane of the sky with time when only the Galactic potential is considered, although the association is found to be moving towards the Galactic center (Fig~\ref{usvs}).
We estimated the gravitational force on a given star due to the total mass of the association and compared it to the force exerted on the star by the Galactic potential. It was found that the binding force due to the association on a given member is highly insignificant when compared with the force due to the Galactic potential, rendering the association as an unbound entity that is slowly expanding with time.

\section{Discussion}
\label{discuss}

The results obtained from \textit{Gaia} DR2 for the parallax, proper motion and peculiar velocities of \cygx{} and \ob{} are consistent with those obtained using the Hipparcos mission. This further supports the identification of \ob{} as the parent association of \cygx{}. We find that, although the parallaxes obtained with \textit{Gaia} are the most accurate and precise measurements available so far, the measurement uncertainties are large enough to introduce scatter in the membership determination of individual stars, as well as a measurement of the physical scale of the association along the line of sight. 

We studied the association \ob{} using the full astrometric solution provided by \textit{Gaia} DR2 and radial velocities obtained from the literature. The distribution of distances (Fig~\ref{dist_histo}) shows that the majority of the stars lie at a distance of 1.92\er{}0.31 kpc.

The star HD~227018 [{\tt 33}] shows the highest proper motion and HD~227943 [{\tt 2}], HD~228041 [{\tt 4}], HD~228104 [{\tt 7}] and HD~227757 [{\tt 28}] are the stars with the smallest proper motions in the sample (Fig~\ref{pm}).
Peculiar velocities could be calculated only for HD~227018 [{\tt 33}] and HD~227757 [{\tt 28}]. The study of \vpec{} values further corroborates the cohesive nature of \ob{} as the majority of stars have peculiar velocities of about 22 \kms{}, with a small velocity dispersion. A comparison of \vpec{} and \vrel{} of individual stars in the association has provided important evidence about the kinematics of stars in the association, and helped to identify those sources with peculiar kinematic behavior. Some of the stars identified in this way are discussed in detail below.

\subsection{Cyg X-1 / HD~226868 [{\tt 33}]}
The black hole binary \cygx{} is located at the boundary of \ob{}. It is important to identify its parent association in order to understand the formation of black holes and natal kicks \citep[see][]{blaauw61, lyne94, vanparad95, white96, hansen97, jonker04, repetto12} imparted during supernovae explosions. Since a majority of known black hole binaries host low mass companions that can be long lived, they could have travelled large distances from their parent associations or site of birth. Therefore it is difficult to identify the parent association of a system with an evolved star and high peculiar velocity. \cygx{} is an important black hole in this regard as it is a nearby and young binary with a massive companion and it has been found to exhibit low peculiar velocities \citep{mirabel-rodriguez-2003, mirabel17-review, mirabel17}. 

It has been suggested by \citet{mirabel17-review} that \cygx{} is at the same distance as the stellar association \ob{} using data from Hipparcos and VLBI radio observations. There are three distance estimates for \cygx{} from geometric parallaxes available in the literature. Hipparcos' new reduction places the source at a distance of 0.60~$\pm$~0.32 kpc, while the distance measured with VLBI radio observations is 1.86~$^{+0.12}_{-0.11}$ kpc \citep{reid11}. The latest distance estimate using geometric parallax measurements from \textit{Gaia} DR2 is 2.37~$\pm$~0.18 kpc \citep{gandhi19}. \cite{massey95} measure the distance to the black hole binary to be 2.14~$\pm$~0.07 kpc using photometric and geometric methods. Though more work is needed to understand systematic differences, all of the above mentioned distance estimates, excluding the distance from Hipparcos' new reduction, are in agreement at the 2$\sigma$ level of confidence, and hence are consistent with the inferred distance to the centre of the association. 

The relative velocity of \cygx{} is very small with respect to the association, as has been discussed by \citet{mirabel-rodriguez-2003}. The results for $r$, $\mu$, \vpec{} and \vrel{}, obtained with the astrometric data from \textit{Gaia} DR2, confirm the findings of \citet{mirabel-rodriguez-2003} and support the conjecture that \cygx{} is a member of \ob{}. The implication of this is that it formed in situ without a strong natal kick. \cygx{} is one of few dynamically measured black holes in high mass X-ray binaries, and its low kick is consistent with weak trends of mass dependent natal kicks observed in black hole binaries \citep[e.g.][]{gandhi19}.

\subsection{HD~227943 [{\tt 2}], HD~228041 [{\tt 4}] and HD~228104 [{\tt 7}]}

These three stars show significantly smaller values of $\mu$ as compared to the rest of the sample. All three are located towards the low end of the range of Galactic latitudes within \ob{}. HD~227943 [{\tt 3}] is located at the boundary of the region (Fig~\ref{propermotion}) selected in this work. HD~227943 [{\tt 3}] and HD~228041 [{\tt 5}] are not in the list of stars of \ob{} published by \citet{garmany92}, however HD~228104 [{\tt 8}] is included. The peculiar velocities for all three stars could not be studied as their radial velocities are not available. Assuming $v_r$ is equal to the median value of the association, the \vpec{} and \vrel{} of the three stars are $\sim$22 \kms{} and $\sim$38 \kms{}, respectively.

\subsection{HD 227757 [{\tt 28}]}
HD 227757 [{\tt 28}] shows a smaller magnitude of proper motion relative to the mean of the association, as shown in Fig~\ref{pm}. In addition, the direction of proper motion differs in Galactic latitude from the rest of the stars in the sample as shown by the blue arrow in Fig~\ref{propermotion}. The estimated \vpec{} of the object is small ($\sim$28 \kms{}), and therefore the object cannot be classified as a runaway star \citep[\vpec{} >30 \kms{}; ][]{blaauw61}. \citet{xu18} measure its velocity with respect to the local standard of rest to be 22.9~\er{}~20.0 \kms{}. \citet{mahy2013} found no significant variation in the radial velocity measurements for this star from spectroscopic studies and suggested the star as `presumably' single. A different direction of proper motion from the majority of the stars in the sample may suggest a possible interaction or encounter in the past. This is further supported by the higher \vrel{} = 43.8 $\pm$ 4.7 \kms{} (see Fig~\ref{vpec}).

\subsection{HD~227018 [{\tt 33}]}
HD~227018 has manifested its distinct kinematic behavior in all diagnostics studied in this work. The highest proper motion of this star amongst our sample is evident from Fig~\ref{propermotion}. It shows \vpec{} of $\sim$50 \kms{} (Fig~\ref{vpec}) and it is a known runaway star moving with a velocity of $>$30 \kms{} \citep{blaauw61, vanburen}. Its relative velocity \vrel{} is also higher than for the other stars ($\sim$40 \kms{}). \citet{tetz11} measure a peculiar spatial velocity of 37.4$^{+4.8}_{-7.2}$ \kms{} and classify the source as a runaway star using Hipparcos data. The star was not included in Cyg OB3 (Cyg 3A and Cyg 3B) by \citet{melnik95}, but was considered as a member of NGC 6871, the core of the Cyg OB3 association, by \citet{massey95}. As discussed in \citet{gvar09} and  \citet{gvar11}, the high velocity of runaway stars can be a consequence of either the disruption of a massive binary following a supernova explosion \citep{blaauw61, stone91, leonard93, iben96} or dynamical three or four-body encounters in dense stellar systems \citep{poveda67, aarseth74, gies86, leonard90,leigh11,leigh12,leigh16,leigh18,ryu17a,ryu17b}. Our analysis reveals that the star HD 227611 [{\tt 23}] came to a distance of closest approach of $\sim$20 pc about 1 Myr ago, however this result in not conclusive of a possible interaction between the two stars resulting in the high velocity of the runaway star.

\section{Summary}
Astrometric data from \textit{Gaia} DR2 for 41 stars located in the region covered by \ob{} are studied in this work. The stars in the sample are found to be concentrated at a heliocentric distance of about 2 kpc.
Most of the stars are observed to be moving with peculiar velocity, \vpec{} of about 22 \kms{}. 
One kinematic subgroup has been identified consisting of majority of the sample stars. A small dispersion in \vpec{} conforms to the results suggesting that the association forms a coherent structure in velocity space \citep[e.g.][]{brown99, tian96, math86}. OB associations are unbound systems formed from single molecular clouds, wherein the stars begin to drift apart after the escape of dust and gas \citep{hum78, melnik17}. We have calculated the relative velocities of individual stars with respect to the association, which can be considered as a measure of this drift. The small relative velocities of $<$ 20 \kms{} for a majority of the stars suggest that \ob{} is a slowly expanding association. Since the relative velocities of the stars are very close to or only marginally different from their respective peculiar velocities, we conclude that expansion is a significant contributor to the peculiar velocities of individual stars.

The kinematic characteristics of \cygx{} are found to be consistent with those of the majority of stars in the sample. The membership of \cygx{} in the association is not ruled out, despite the location of the binary at the boundary of the association. This result is in agreement with \citet{mirabel17-review}, \citet{mirabel17}, and \citet{mirabel-rodriguez-2003} that \cygx{} is most likely a member of the association and was formed in situ. 

Our analysis reveals interesting kinematic behavior for some of the stars; e.g. HD 227757 [{\tt 28}], which exhibits a proper motion in a direction different to the majority of the stars in the association. HD 191611 [{\tt 27}] is found to have \vrel{} $>$ \vpec{}. The peculiar kinematic properties of these stars might be remnants of past encounters of the stars in a dense environment. The status of HD 227018 as a runaway star is further supported by data from \textit{Gaia} DR2. The rest of the stars in the sample display \vpec{} $\lesssim$ 30 km s$^{-1}$ and are not classified as runaway stars.  

Finally, projected positions of the sample stars on the Galactic xy-plane reveals an extended structure along the line of sight.
This is likely an artifact of the uncertainties in the parallax measurements. At the present level of uncertainties, we restricted ourselves to the study of the kinematic properties of the stars and refrained from addressing spatial structure and from identifying members and non-members in the association. Future \textit{Gaia} data releases with improved uncertainties will make such studies more feasible.

\section*{Acknowledgment}
The authors thank the anonymous reviewer for useful comments that helped to improve the paper. This work has made use of data from the European Space Agency (ESA) mission {\it Gaia} (\url{https://www.cosmos.esa.int/gaia}), processed by the {\it Gaia} Data Processing and Analysis Consortium (DPAC, \url{https://www.cosmos.esa.int/web/gaia/dpac/consortium}). Funding for the DPAC has been provided by national institutions, in particular the institutions participating in the {\it Gaia} Multilateral Agreement. AR acknowledges a Commonwealth Rutherford Fellowship. JP is in part supported by funding from a University of Southampton Central VC Scholarship. PG thanks STFC for support (ST/R000506/1). NWCL gratefully acknowledges a Fondecyt Iniciac\'ion grant (\#11180005).  DB thanks Magdalen College for his fellowship and the Rudolf Peierls Centre for Theoretical Physics for providing office space and travel funds. This work is supported by a UGC-UKIERI Phase 3 Thematic Partnership. This research has made use of observatory archival image from DSS2.


\label{lastpage}
\end{document}